\documentclass[prl,twocolumn,showpacs,amsmath,amssymb,superscriptaddress]{revtex4}
\usepackage{graphicx}
\usepackage{dcolumn}
\usepackage{bm}
\usepackage{epsfig}

\begin{document}
\title{What relaxes muon spins in molecular nanomagnets?}
\author{T. Lancaster}
\email{t.lancaster1@physics.ox.ac.uk}
\author{S.J. Blundell} 
\affiliation{Clarendon Laboratory, Oxford University Department of Physics, Parks
Road, Oxford, OX1 3PU, UK}
\author{F.L. Pratt}
\affiliation{ISIS Facility, Rutherford Appleton Laboratory, Chilton, 
Oxfordshire OX11 0QX, UK}
\author{I. Franke}
\affiliation{Clarendon Laboratory, Oxford University Department of Physics, Parks
Road, Oxford, OX1 3PU, UK}
\author{A.J. Steele}
\affiliation{Clarendon Laboratory, Oxford University Department of Physics, Parks
Road, Oxford, OX1 3PU, UK}
\author{P.J. Baker}
\affiliation{ISIS Facility, Rutherford Appleton Laboratory, Chilton, 
Oxfordshire OX11 0QX, UK}
\author{Z. Salman}
\affiliation{Laboratory for Muon-Spin Spectroscopy, Paul Scherrer Institut, Villigen, CH}
\author{C. Baines}
\affiliation{Laboratory for Muon-Spin Spectroscopy, Paul Scherrer Institut, Villigen, CH}
\author{I. Watanabe}
\affiliation{Muon Science Laboratory, RIKEN, 2-1 Hirosawa, Wako, Saitama 351-0198, Japan}
\author{S. Carretta}
\affiliation{Dipartmento di Fisica, Universit\`{a} di Parma, I-43100 Parma, Italy}
\author{G.A. Timco}
\affiliation{Department of Chemistry, University of Manchester, Oxford Road, Manchester,
M13 9PL, UK}
\author{R.E.P. Winpenny}
\affiliation{Department of Chemistry, University of Manchester, Oxford Road, Manchester,
M13 9PL, UK}

\begin{abstract}
We address the cause of the unusual muon spin relaxation ($\mu^{+}$SR) results on molecular nanomagnets (MNMs).
Through measurements on protonated and deuterated
samples of the MNMs Cr$_{7}$Mn ($S=1$) and Cr$_{8}$ ($S=0$),
we show that the muon spin for $S \neq 0$ MNMs is relaxed via dynamic fluctuations of the electronic
spins. A freezing out of dynamic processes occurs on cooling and at low temperatures
the muon spins are relaxed by the electronic spins which themselves are dephased by 
incoherent nuclear field fluctuations.
We observe a transition to a state of static magnetic order of the MNM electronic spins
in Cr$_{7}$Mn below 2~K. 
\end{abstract}
\maketitle

\begin{figure}
\begin{center}
\epsfig{file=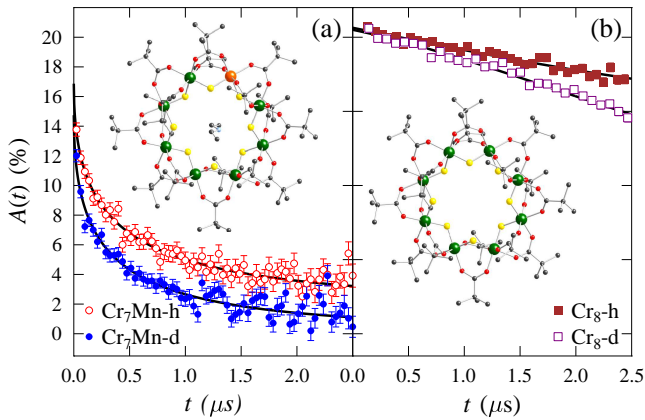,width=\columnwidth}
\caption{(Color online.)  $\mu^{+}$SR spectra for protonated and deuterated
(a) Cr$_{7}$Mn and (b) Cr$_{8}$ materials, measured at $T=4.5$~K.
Inset: structures of the molecules.
\label{data}}
\end{center}
\end{figure}

Molecular nanomagnets (MNMs)\cite{gatteschi} comprise clusters of exchange-coupled
transition metal ions which often have a negative anisotropy constant favoring
a ground state with a large eigenvalue of the electronic spin component $S_{z}$.
These systems have been widely studied in
recent years, most recently in anticipation of their deployment as elements of
quantum computers \cite{loss,ardavan}, although much interest also centers on
the quantum tunnelling of the magnetization
(QTM) which takes place when the magnetic energy levels are at resonance 
\cite{gatteschi}. 
The theory of QTM is based on the Landau--Zener paradigm \cite{landau} of 
energy levels brought into resonance by a time dependent field. 
The precise details of the mechanism for QTM in MNMs remain unresolved and
while pairwise dipole interactions between spins are certainly important \cite{morello}
a
promising suggestion is that resonance is achieved as a result of the
stochastically varying magnetic field  $\boldsymbol{B}_{\mathrm{n}}(t)$, which
arises from the incoherent fluctuations of nuclear moments
surrounding the transition metal ions \cite{stamp} (a typical
MNM contains $\sim 10^{2}$ protons due to the organic ligands).

MNMs have been  successfully probed using techniques such as 
magnetization \cite{cornia}, heat capacity \cite{hc}, neutron scattering \cite{ns,caciuffo} 
and electron spin resonance \cite{ardavan}. 
In contrast, measurements made using 
muon spin relaxation ($\mu^{+}$SR) have proven difficult to interpret. Although initially
it was thought that QTM should be measurable by implanting muons into 
MNMs \cite{lascialfari,salman,blundell}, 
the unambiguous detection of 
this effect proved elusive \cite{lancaster}. 
Instead, $\mu^{+}$SR spectra obtained on high spin systems appeared to arise
 from dynamic
fluctuations of a local magnetic field distribution at the muon sites, which persisted 
down to dilution refrigerator temperatures \cite{salman,blundell,lancaster}.
Muon results on MNM systems all showed similar behavior \cite{lancaster} but
 it was unclear whether the muon was probing the intrinsic behavior of the
large electronic spin or some residual effect.
It was recently argued by Keren {\it et al.} \cite{keren} that $\mu^{+}$SR is
sensitive to the dephasing of the 
MNM electronic spins 
caused by the incoherent fluctuations of nuclear moments in which the metal ions are embedded. 
If this is the case then it makes the muon a valuable probe of the potential mechanism behind
QTM.
In order to address the question of what the muon probes in MNM systems
 we have made identical $\mu^{+}$SR measurements on
protonated and deuterated samples of  Cr$_{7}$Mn ($S=1$) and Cr$_{8}$ ($S=0$) \cite{larsen,slagaren,footnote}
(structure shown in Fig.~\ref{data}). 
Here we show (i) that the muon is controlled by the large electronic spin in an MNM;
(ii)  deuteration
leads to a significant increase in the $\mu^{+}$SR relaxation rate at low temperature in Cr$_{7}$Mn,
implying that muons probe the dephasing of large electronic spins
by the random magnetic fields due to the nuclei and
(iii) that upon cooling, a magnetic ground state is reached by a
freezing out of dynamic processes that leads to magnetic order in Cr$_{7}$Mn below 2~K.

In a $\mu^{+}$SR experiment \cite{steve}, spin-polarized
positive muons are stopped in a target sample.
The time evolution of the muon spin polarization is probed via the
positron decay asymmetry function $A(t)$ to which it is proportional.
Our $\mu^{+}$SR measurements were made on the 
ARGUS instrument at the RIKEN-RAL facility, ISIS, Rutherford Appleton Laboratory, UK
and on the 
LTF and GPS instruments at the Swiss Muon
Source (S$\mu$S), Paul Scherrer Institut, CH. 
Powder samples of the materials were packed in
Ag foil and mounted on a Ag backing plate in
$^{3}$He and $^{4}$He cryostats. 
Typical spectra measured for Cr$_{7}$Mn and Cr$_{8}$
are shown in Fig.~\ref{data}. Above $T\approx 2$~K
the spectra for all materials differ depending on
whether protonated or deuterated (Fig.~\ref{data}). 

To compare the Cr$_{7}$Mn and Cr$_{8}$ systems, measurements 
were made over the temperature range $2 \leq T \leq 100$~K.
In this regime the spectra for $S=1$ Cr$_{7}$Mn (Fig.~\ref{data}(a)) were found to be
described by the relaxation function
$
A(t) = A_{1} \exp(-\sqrt{\lambda t}) + A_{\mathrm{bg}},
$
where $A_{\mathrm{bg}}$ accounts for any background contribution from
muons that stop in the sample holder or cryostat tails. 
This 
behavior is typical of that observed previously in 
MNM materials \cite{salman,blundell,keren} and 
arises because of the complex dynamic distribution of local fields within the material 
sampled by the muon ensemble. 
The monotonic relaxation and the fact that the muons
could not be decoupled with an  applied magnetic field up to 0.6~T places the
relaxation in the fast-fluctuation limit \cite{hayano}.

\begin{figure}
\begin{center}
\epsfig{file=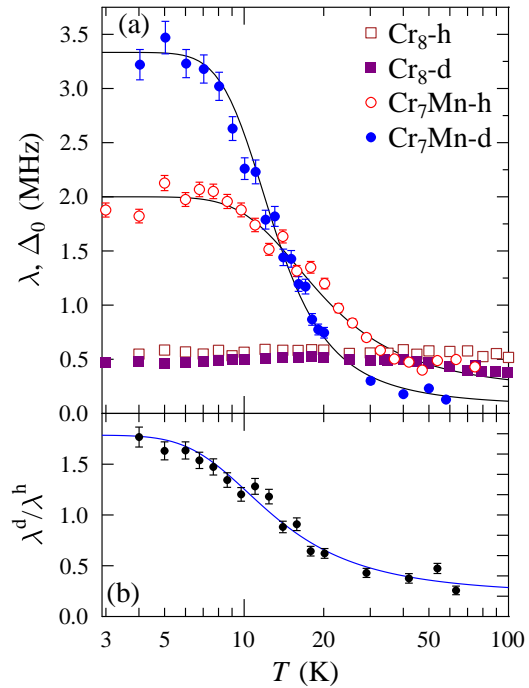,width=7cm}
\caption{(Color online.) (a) Temperature evolution of the relaxation rates $\lambda$
for Cr$_{7}$Mn and $\Delta_{0}$ for Cr$_{8}$. The lines are fits to a phenomenological
equation (see main text). 
(b) Ratio of the Cr$_{7}$Mn-h and -d  relaxation rates. 
The line is a guide to the eye.
\label{results}}
\end{center}
\end{figure}

The spectra measured for the $S=0$ Cr$_{8}$ samples are quite different (Fig~\ref{data}(b)). In this case the
relaxation rate is far smaller and resembles a Kubo-Toyabe (KT) function  $f_{\mathrm{KT}}(\Delta t)$
with $\Delta = \gamma_{\mu} \langle B^{2} \rangle$ where $\gamma=2 \pi \times 135.5$~MHz~T$^{-1}$
is the muon gyromagnetic ratio and $B$ is the local magnetic field at a muon site \cite{hayano}.
However, the KT function, which describes relaxation due to static random magnetic fields at the muon sites,
could not adequately describe all of the data. This is probably to be 
expected in a 
complex material such as a MNM where many inequivalent classes of muon sites occur and lead to a distribution
of second moments $p(\Delta)$. The resulting muon relaxation is obtained by averaging the KT function 
over this distribution. For
simplicity we modeled the relaxation by taking $p(\Delta)$ to be a uniform distribution up to a maximum $\Delta_{0}$. 
This one-parameter model successfully fitted the data for both materials over the entire time range. 
The fitted 
values of $\Delta_{0}$ are shown in Fig.~\ref{results}(a) and are quite $T$-independent
and take average values $\Delta_{0}=0.56(1)$~MHz for Cr$_{8}$-h and $0.48(1)$~MHz for Cr$_{8}$-d.
Our conclusion from these measurements is that the muon is sensitive to the  disordered 
nuclear moments in Cr$_{8}$. This is confirmed by the application of a small longitudinal 
magnetic field which quenches the relaxation. 
The larger $\Delta_{0}$ in Cr$_{8}$-d compared to Cr$_{8}$-h reflects (albeit partially) the larger moment of 
the deuteron.
Most importantly, the dramatic difference between the measured spectra and relaxation rates 
for $S=0$ Cr$_{8}$ and $S=1$ Cr$_{7}$Mn samples (Fig.~\ref{data}) strongly suggests that the muon response 
in MNM systems with $S \neq 0$ stems from dynamic
fluctuations of the {\it electronic} spin. In the absence of an electron spin in Cr$_{8}$,
the muon spin is relaxed by quasistatic disordered nuclear moments.

We now turn to the role of the nuclei in the spin relaxation process in $S = 1$ MNMs. 
The temperature dependence of the relaxation rate $\lambda$ for the 
protonated ($\lambda^{\mathrm{h}}$) and deuterated ($\lambda^{\mathrm{d}}$) Cr$_{7}$Mn samples
 is shown in Fig.~\ref{results}(a).  
These measurements were made with a magnetic field of 1~mT applied parallel to the initial
muon spin direction which was intended to quench any residual relaxation of the muon spin 
caused directly by fluctuating nuclear spins as seen in the Cr$_{8}$ samples.
On cooling below $T \sim 50$~K, the relaxation rate $\lambda$ increases before saturating below $\sim 10$~K. 
with the onset of the increase and the saturation
occurring at similar values of $T$ for both materials.
This $T$-dependence is common to nearly all MNM
systems that have been previously measured with $\mu^{+}$SR 
\cite{salman,blundell,lancaster,keren} and is discussed in more detail below.
At high $T$ we see that $\lambda^{\mathrm{d}} > \lambda^{\mathrm{h}}$.
It is likely that at these high temperatures
the electronic spins are fluctuating very fast and are at least partially motionally narrowed from the
spectra.
Upon cooling the increase in $\lambda$ is greater for the deuterated sample, 
with $\lambda^{\mathrm{d}}$  becoming greater than $\lambda^{\mathrm{h}}$ below $\approx 15$~K. 
Most significantly the saturation of the relaxation at $T \lesssim 10$~K 
occurs with $\lambda^{\mathrm{d}} > \lambda^{\mathrm{h}}$. The $T$-dependence of the ratio
$\lambda^{\mathrm{d}}/\lambda^{\mathrm{h}}$ [Fig.~\ref{results}(b)] which increases on cooling
tends to $\approx 1.7$ at the lowest temperature.

The low-temperature results may be understood using the model of $\mu^{+}$SR in MNM systems
proposed by Keren {\it et al.} \cite{keren}. Here a muon spin $\boldsymbol{I}$ is coupled to a single MNM 
electronic 
spin $\boldsymbol{S}$ and 
dynamic fluctuations of the local protons cause $\boldsymbol{S}$ to
experience a nuclear stochastic field $\boldsymbol{B}_{\mathrm{n}}(t)$, along with any externally applied field 
$\boldsymbol{B}_{\mathrm{a}}$. 
This system may be described by the resulting Hamiltonian
$
H = g_{\mathrm{s}} \mu_{\mathrm{B}} \boldsymbol{S} \cdot [\boldsymbol{B}_{\mathrm{a}} + \boldsymbol{B}_{\mathrm{n}}(t)]
+ \hbar \gamma_{\mu}  \boldsymbol{I} \cdot [\boldsymbol{B}_{\mathrm{a}} + \mathbf{F} \boldsymbol{S}],
\label{hamiltonian}
$
where $\mathbf{F}$ couples the muon and
electronic spin. We ignore the direct response of the muon to the stochastic field, 
which we know to be small from the Cr$_{8}$ results. 
The spin relaxation resulting from this model may be simply understood. The magnetic
field distribution experienced by a particular spin population is characterized by a correlation
function whose Fourier transform (known as the spectral weight) will be proportional to the 
spin relaxation rate. 
We assume that the stochastic magnetic field  $\boldsymbol{B}_{\mathrm{n}}(t)$ experienced by the MNM spin
due to the nuclei will be described by
$
\langle \boldsymbol{B}_{\mathrm{n}}(t) \boldsymbol{B}_{\mathrm{n}}(0) \rangle 
=  \langle {B_{n}}^{2}\rangle \exp(-t/\tau_{\mathrm{n}}),
$
where $\tau_{\mathrm{n}}$ is the correlation time of the nuclear stochastic field. A Fourier transform 
reveals that the nuclear fields will relax the electron spins at a rate \cite{keren}
$
1 / \tau_{\mathrm{e}} \propto \langle B_{\mathrm{n}}^{2} \rangle \tau_{\mathrm{n}}.
\label{taue}
$
 To find the relaxation rate of the muon spin due to the fluctuations
of the electronic spins, we again assume that the correlation function takes the form
$
\langle \boldsymbol{B}_{\mu}(t) \boldsymbol{B}_{\mu}(0) \rangle =  \langle B_{\mu}^{2} \rangle \exp(-t/\tau_{\mathrm{e}}),
$
where $\boldsymbol{B}_{\mu}$ is the effective local field at the muon site due to coupling to the large spin $S$
and $\tau_{\mathrm{e}}$ is the correlation time for the electronic spins. 
Taking a further Fourier transform gives the $\mu^{+}$SR relaxation rate  \cite{hayano,keren} as
$
\lambda = \gamma_{\mu}^{2} \langle B_{\mu}^{2} \rangle  \tau_{\mathrm{e}}/
[1 + (\gamma_{\mu} |B_{\mathrm{a}}| \tau_{\mathrm{e}})^{2}]$
which in the limit of small applied magnetic fields becomes
\begin{equation}
\lambda = \gamma_{\mu}^{2} \langle B_{\mu}^{2} \rangle  \tau_{\mathrm{e}} \propto 
\frac{
\langle B_{\mu}^{2} \rangle}{\langle B_{\mathrm{n}}^{2} \rangle}\frac{1}{ \tau_{\mathrm{n}} }.
\label{muonlambda}
\end{equation}

The primary effect of swapping protons for deuterons might be expected to be the factor 3.26
decrease in the size of the nuclear moments
experienced by the electronic spin and this should be expected to reduce the second moment of the field 
distribution due to the electronic spin $\langle B_{\mu}^{2} \rangle$.
In fact there is good evidence for this from ESR \cite{ardavan} in the
similar system Cr$_{7}$Ni, 
where it was found that $1/T_{2}$ decreased by the expected factor of 6 (corresponding to the
reduction in nuclear gyromagnetic ratios) upon deuteration,
demonstrating that the electronic spins are directly relaxed by proton fluctuations. 
Eq.~(\ref{muonlambda}) shows that a decrease in $\langle B_{\mathrm{n}}^{2} \rangle$
 will lead to an {\it increase} in the muon relaxation rate $\lambda$. The
larger magnitude of $\lambda$ for the Cr$_{7}$Mn-d sample at low temperatures  
is therefore consistent with  the electronic spin being dephased by the nuclei. 
Since the electronic spins, and hence $\langle B^{2}_{\mu} \rangle$, 
have the same magnitude in both materials, then we have
$\lambda^{\mathrm{d}} / \lambda^{\mathrm{h}} =  
\langle B_{\mathrm{n}=\mathrm{h}}^{2} \rangle \tau_{\mathrm{n}=\mathrm{h}} / 
\langle B_{\mathrm{n}=\mathrm{d}}^{2} \rangle \tau_{\mathrm{n}=\mathrm{d}}$.
We note that the measured increase at low temperatures is, at most, a factor of $\approx 1.7$, 
rather than a larger 
factor that might be predicted if the distribution width scales with the moment sizes alone. 
It is probable that swapping protons for deuterons changes not only 
the local field distribution but also its correlation
time. 

Finally we address the characteristic temperature dependence of $\lambda$
observed in all $S \neq 0$ MNM systems.
It is generally found in MNMs \cite{lancaster} that on decreasing the temperature from $T \sim 100$~K 
there is an increase in $\lambda$ with the relaxation rate
levelling off to some value $\lambda_{\mathrm{sat}}$ below $\sim 10$~K. 
This behavior may be explained by involving two competing dynamic relaxation processes, one dominant
at high temperatures and 
described by a strongly temperature-dependent correlation time $\tau_{s}(T)$ and 
one, dominant at low temperatures described by a weakly $T$-dependent correlation time $\tau_{\mathrm{w}}$.
 In the presence of two competing processes, that with the shorter correlation
time dominates, giving the smaller relaxation rate (since $\lambda \propto 1/\tau$). 
At high temperature, therefore, we have $\tau_{\mathrm{s}}(T) \ll \tau_{\mathrm{w}}$ which results 
in a
strongly $T$ dependent relaxation which we can crudely model phenomenologically with $\lambda = C  \exp (U / T)$. 
At low temperatures when some of the $T$-dependent relaxation channels have been frozen out, 
$\tau_{\mathrm{w}} \ll \tau_{\mathrm{s}}(T)$ and
we have $\lambda \sim \lambda_{\mathrm{sat}}$. This behavior
results in a phenomenological fitting function 
$1/ \lambda(T) = 1/\lambda_{\mathrm{sat}} + 1/[C \exp (U/T)]$ which has been used previously
to characterize these MNM systems \cite{lancaster}. Fitting this formula to our data 
(Fig.~\ref{results}(a)) yields $C^{\mathrm{h}}=0.23(3)$~MHz, $U^{\mathrm{h}} = 46(4)$~K, 
$\lambda^{\mathrm{h}}_{\mathrm{sat}}=2.00(4)$~MHz, $C^{\mathrm{d}}=0.070(1)$~MHz,
 $U^{\mathrm{d}} = 51(3)$~K and $\lambda^{\mathrm{d}}_{\mathrm{sat}}=3.3(1)$~MHz. 

There is a resemblance between the muon results in
MNM systems and those in some inorganic materials, such as
Ca$_{3}$CoMnO$_{6}$ \cite{tom}, 
where the physics involves significant single-ion anisotropy and
a complex (often glassy) freezing out of dynamic processes.
It is probable that the dephasing of the electronic spins we probe in MNMs is explainable within the same framework. 
At high temperatures magnetoelastic interactions
provide the main relaxation mechanism for the 
electronic degrees of freedom.
Our data show $U^{\mathrm{h}} \approx U^{\mathrm{d}}$
since, if the $T$-dependence is due to
spin-phonon coupling through the modulation of local crystal fields, 
this barrier height should only depend on the electronic energy level structure. 
As the temperature is decreased, some relaxation channels will be frozen out, increasing the
correlation time of the electronic moments \cite{nmr}. The low temperature channel that
gives rise to the temperature independent relaxation 
$\lambda^{\mathrm{sat}}$ seen in $\mu^{+}$SR would appear to be the 
relaxation of the electronic spins by the nuclear fluctuations, allowing us to identify
$\tau_{\mathrm{w}}$ with $\tau_{\mathrm{e}}$, the electronic correlation time
discussed above. 
This situation is similar to the electronic $T_{2}$
which is dominated by phononic contributions at high
temperatures and by nuclear contributions at low temperature \cite{troiani}. 
This is not the case for the electronic
$1/T_{1}$ which is phonon dominated down to $\sim 2$~K. 
This difference between $1/T_{1}$ and $1/T_{2}$ for ESR suggests that the
nuclei contribute to the secular part of the relaxation, that is, the dephasing of the
electron spins due to a spread in the net magnetic field at the spin sites \cite{ardavan,slichter}.

\begin{figure}
\begin{center}
\epsfig{file=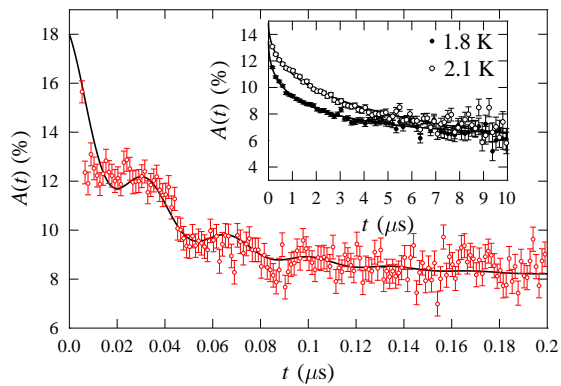,width=7.6cm}
\caption{(Color online.) 
Example spectrum measured at 0.5~K for Cr$_{7}$Mn-h at (S$\mu^{+}$S) showing heavily damped oscillations with
a frequency of $\gamma_{\mu} B/(2 \pi) \sim 30$~MHz.
{\it Inset:} spectra from ISIS show a discontinuous change on cooling through below 2~K. 
\label{data2}}
\end{center}
\end{figure}

In order to further probe the freezing out of the dynamics in Cr$_{7}$Mn,
measurements were made down to 20~mK using the LTF instrument at S$\mu$S. 
At the lowest temperatures heavily damped oscillations are observed at early times (Fig.~\ref{data2})
in both Cr$_{7}$Mn-h and -d. 
Oscillations are usually caused in muon spectra by quasistatic magnetic order, 
causing a coherent precession of muon spins. The observed oscillations show little temperature 
dependence in the
range $0.02 \leq T \leq 1$~K and are identical for -h and -d samples.
For measurements made in the same temperature range at ISIS, oscillations are not discernible due to the
resolution limit set by the ISIS muon pulse-width. However, the spectra do show a discontinuous 
change around $T=2$~K [inset Fig.~\ref{data2}], 
with a loss of initial asymmetry and a sharp increase in the apparent relaxation rate observed below 
this temperature. There is also significant magnetic hysteresis on the application and removal
of applied magnetic fields below 2~K.
From these measurements we estimate that a transition to a state of increased static magnetic order
takes place at $T=1.9(1)$~K. The heavily damped nature of the oscillations and the Cr$_{8}$ results
suggest that there 
are many magnetically inequivalent muon sites in the system. The results demonstrate that the
intermolecular exchange $J$ is nonzero in the $S=1$ system and, using mean field theory \cite{blundell2}
(assuming $z=6$ nearest neighbors)
we estimate $J/k_{\mathrm{B}} = 3 T_{\mathrm{C}}/2 z S(S+1) \approx 0.2$~K.

This work was carried out at S$\mu$S
and the RIKEN-RAL facility (ISIS)  supported by a beamtime allocation from the 
Science and Technology Facilities Council.
We are grateful to Alex Amato for experimental assistance and the EPSRC (UK)
for financial support.

\end{document}